\documentclass[apl,reprint,showpacs,twocolumn,superscriptaddress]{revtex4-1}
\pdfoutput=1

\usepackage[colorlinks,linkcolor=blue,citecolor=blue]{hyperref}
\usepackage{graphicx}
\usepackage{natbib}
\usepackage{comment}
\usepackage{amsmath}
\usepackage{amssymb}
\usepackage{amsfonts}
\usepackage{upgreek}

\begin{document}


\title{MicroSQUIDs based on V/Cu/V Josephson nanojunctions}

\author{Alberto Ronzani}
\email[]{alberto.ronzani@sns.it}
\affiliation{NEST, Istituto Nanoscienze-CNR and Scuola Normale Superiore,
             I-56127 Pisa, Italy}

\author{Matthieu Baillergeau} \affiliation{D\'epartement de Physique, Ecole Normale Sup\'erieure, 
             24 Rue Lhomond, F-75005 Paris, France}

\author{Carles Altimiras}
\affiliation{NEST, Istituto Nanoscienze-CNR and Scuola Normale Superiore,
             I-56127 Pisa, Italy}

\author{Francesco Giazotto}
\affiliation{NEST, Istituto Nanoscienze-CNR and Scuola Normale Superiore,
             I-56127 Pisa, Italy}


\begin{abstract}
	We report on the fabrication and characterization of microSQUID devices 
	based on nanoscale vanadium/copper/vanadium Josephson weak links.
	Magnetically driven quantum interference patterns have been measured for 
	temperatures in the $0.24 - 2 \, \mathrm{K}$ range.
	As magnetometers, these devices obtain flux-to-voltage
	transfer function values as high as $450 \, \upmu \mathrm{V}/\Phi_0$
	leading to promising magnetic flux resolution  
	$\Phi_{\mathrm{N}} < 3 \, \upmu \Phi_0 / \sqrt{\mathrm{Hz}}$, 
	being here limited by the room temperature
	preamplification stage. Significant improvements in the flux
	noise performance figures, which are already competitive with
	existing state-of-the-art commercial SQUIDs systems, are expected either with
	cryogenic preamplification or with the adoption 
	of optimized electronic readout stages based on active feedback schemes.
\end{abstract}



\maketitle



A superconducting quantum interference device (SQUID) is a magnetic flux sensor
constituted by a parallel circuit of two superconducting
weak links forming a ring. An external magnetic field threading this loop  
controls the electron transport properties of the Josephson weak links
via flux quantization\cite{Deaver1961,Doll1961} and the DC Josephson 
effect\cite{Josephson1962} therefore modulating the total amount 
of supercurrent flowing through the circuit.
SQUIDs based on low critical temperature superconductors realize extremely sensitive
magnetic and current detectors, able to reach nowadays flux noise levels in the 
$\sim 0.3-5 \, \upmu \Phi_0 / \sqrt{\mathrm{Hz}}$ range at liquid He 
temperature\cite{Braginski2004},
with immediate applications in biomagnetism,
nuclear magnetic resonance and susceptometry, 
investigation on the magnetic properties of 
small spin populations\cite{Granata2010,Tilbrook2009},
as well as low-noise readout 
stages for microbolometer detectors\cite{Giazotto2008}.

\begin{figure}[t!]
\includegraphics[width=\columnwidth]{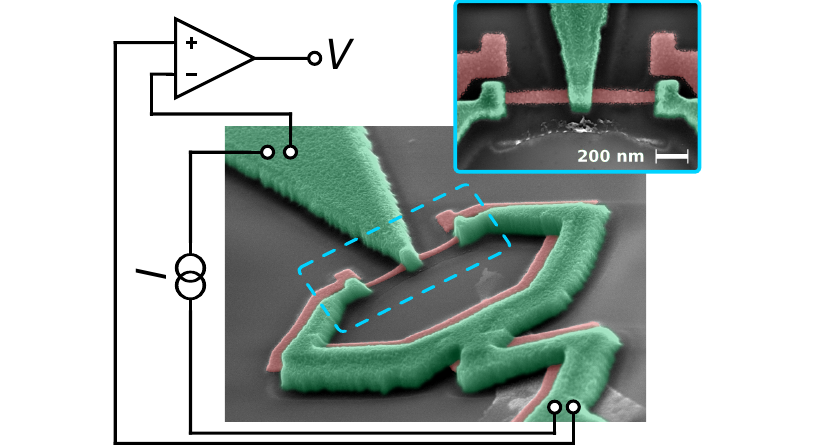}
\includegraphics{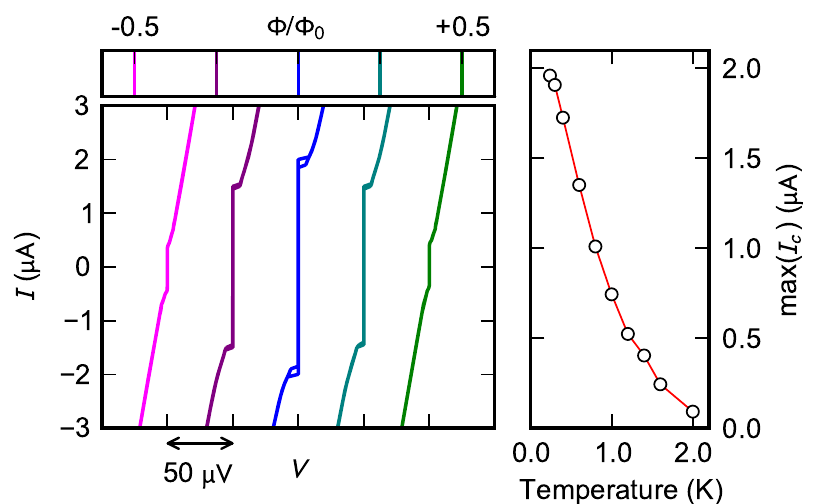}
\caption{
\textsc{Top:} 
Tilted scanning electron micrography showing a typical SQUID device in 
pseudocolors. Vanadium (green) and copper (red) films are $150$ and $20\, \mathrm{nm}$ 
thick,
respectively. The standard setup for a four wire measurement is also displayed as a
superimposed scheme. The inset in the top right corner shows a perpendicular blow-up
of the weak links.
\textsc{Bottom left:}
Current vs voltage ($I$-$V$) characteristics of device A measured at $240 \,\mathrm{mK}$ 
for increasing values of perpendicular magnetic field.
The curves have been horizontally offset for clarity.
\textsc{Bottom right:}
Temperature dependence of the maximum critical current for device A.
The line is a guide to the eye.
\label{fig:sem}}
\end{figure}

\begin{figure*}
 \includegraphics{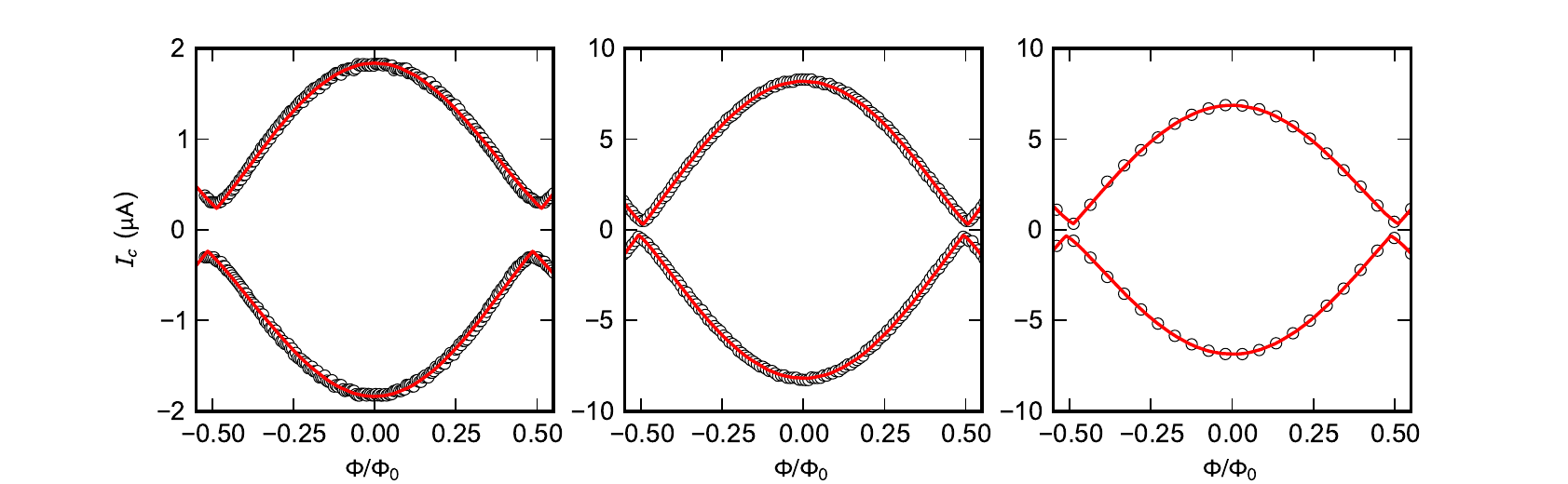}
 \caption{
 $I_{\mathrm{c}}(\Phi)$ dependence for A,B,C devices (left to right).
 Values for the critical current $I_{\mathrm{c}}$ have been extracted from the switching
 current points in the $I$-$V$ characteristics measured at $240 \, \mathrm{mK}$.
 Data points (black circles with diameter corresponding to 
 the experimental uncertainty) have
 been fitted (red line) with the theoretical model displayed in 
 equations~(\ref{eqn:modelpre}, \ref{eqn:model1}). 
 Values for the fitting parameters are reported in table~\ref{tab:samples}.
 \label{fig:icfit}}
\end{figure*}

The vast majority of SQUID systems are based on 
superconductor/insulator/superconductor (SIS) weak links, yet the DC
Josephson effect can also be observed in a number of superconducting 
systems\cite{Likharev1979},
such as constrictions between two superconductor 
banks\cite{Levenson-Falk2013,Vijay2010a}, 
or weak links constituted by normal metal\cite{Savin2004}
or semiconductor elements\cite{Giazotto2011a,Carillo2006,Giazotto2004}. 
Superconductor/normal~metal/superconductor (SNS) junctions
are able to carry a dissipationless phase-dependent supercurrent via the 
\textit{proximity effect}\cite{Pannetier2000}.
The latter induces superconducting correlations in the electronic states of the normal
metal when it is placed in good electric contact with a superconductor.
Such correlations follow from the building of Andreev bound states in the 
N~region\cite{Andreev1964,McMillan1968,Belzig1999}.

The interest in SNS Josephson junctions is justified by their simple, accessible
and reproducible fabrication process. On the one side, the current-phase
relation of SNS weak links can be tailored and controlled beyond the conventional
sinusoidal shape\cite{Heikkila2002} typical of SIS junctions. 
This way one can obtain sharper phase responsivity or have access to non-trivial
states such as the $\pi$-shifted Josephson junction\cite{Baselmans1999}.
On the other side, while SIS Josephson junctions are typically characterized by 
a considerable capacitance due to the presence of the thin oxide layer of the barrier
limiting the performance of SQUIDs\cite{Schmelz2012a},
SNS Josephson junctions do not suffer from this drawback thanks to their
natively negligible capacitance.

A previous work\cite{Angers2008a} on SQUID magnetometers based on SNS weak links
reported devices characterized by critical currents with values of tens of
microamperes. However, the usability of these devices was limited by 
the presence of significant hysteretic behaviour in the voltage-current characteristic
curves, a feature typical of high critical current SNS weak links 
due to heating of the electron gas in the normal domain whenever the junction 
switches to the resistive state\cite{Courtois2008a}.

In order to overcome this issue, we fabricated SQUID devices in which the SNS junctions
are somewhat short yet resistive, so that the voltage modulation range (proportional to
$V_{\mathrm{Th}} = \hbar D / (e L^2)$, the Thouless voltage of the normal wire,
where $D$ is the diffusion constant, $e$ is the electron charge and 
$L$ is the length of the wire)
is maximized while at the same time the Joule dissipation in the resistive regime is 
kept as low as possible to quench any thermal hysteresis.
The requirement for obtaining resistive SNS junctions can be met via the realization of
N wires with nanoscale dimensions.

The top panel of Fig.~\ref{fig:sem} shows a scanning electron micrograph of a typical
interferometer, fabricated by
standard electron beam shadow-mask lithography followed by tilted evaporation
of Cu/V ($20/150\, \mathrm{nm}$) in a UHV chamber. 
The width and length of copper nanojunctions
shown in the inset of the top panel of Fig.~\ref{fig:sem} are $60 \, \mathrm{nm}$ and 
$370 \, \mathrm{nm}$, respectively.
The loop of the SQUID spans a surface of $\approx 1.5 \, \upmu \mathrm{m}^2$.
The superconductor of choice, vanadium, shows several attractive features from an
applied point of view, such as a sizeable critical temperature 
($T_{\mathrm{c}} \approx 5.4 \, \mathrm{K}$
for thick films) resulting in a strong proximization capability over copper domains, 
also made possible by the good quality of the interfaces formed between these two 
metals\cite{Garcia2009}.

The electron magneto-transport properties of the SQUIDs were characterized 
in a filtered ${}^3\mathrm{He}$ cryostat having a base
temperature of $\approx 240\,\mathrm{mK}$. Current vs voltage characteristics 
were recorded
via lock-in technique by measuring the first harmonic of the voltage response
to a DC current bias chopped at a reference frequency ($f \approx 15 \, \mathrm{Hz}$).
A $40\, \mathrm{dB}$ room temperature
low-noise voltage preamplifier (NF Corp. model LI-75A) 
has been used to boost the signal level to be fed to the digital lock-in amplifier
(NF Corp. model LI-5640).

The current vs voltage ($I$-$V$) characteristics of device A
for increasing values of magnetic field applied perpendicularly to the
SQUID plane are presented in the bottom left panel of Fig.~\ref{fig:sem}. 
The characteristic curves show a distinct supercurrent branch; the critical current
$I_{\mathrm{c}}$ being the maximum current that can be sustained in this branch, whose 
value is modulated by the magnetic field applied to the loop. A small residual
hysteresis of thermal origin\cite{Courtois2008a} is present in the 
characteristics for which
the critical current $\left| I_{\mathrm{c}} \right| \gtrsim 1.5 \, \upmu \mathrm{A}$.
As the bias current $I$ exceeds $I_{\mathrm{c}}$, the system switches to 
a resistive state developing a potential difference across the superconducting loop. 
For large bias currents, $I \gg I_{\mathrm{c}}$, the characteristic curve 
can be approximated by $ V \approx I R_{\mathrm{n}}/2 $,
where $R_{\mathrm{n}}$ is the normal-state resistance of each weak link.
For each known geometry of the copper wire, 
the measurement of $R_{\mathrm{n}}$ allows to estimate  
the diffusion coefficient and, consequently, the Thouless energy
($E_{\mathrm{Th}} = e V_{\mathrm{Th}}$) of the weak links. 
Table~\ref{tab:samples} summarizes the values of these parameters for the different 
measured devices.
The temperature dependence of the maximum value of the critical current 
(i.e., that at $\Phi = 0$) for device A is
presented in the bottom right panel of Fig.~\ref{fig:sem}. Magnetically modulated 
supercurrent branches have
been measured up to $\approx 2 \, \mathrm{K}$.

\begin{table}
 \begin{ruledtabular}
 \begin{tabular}{ccccccc}
	 Device & $L$/$W$/$t$      & $R_{\mathrm{n}}$            & $E_{\mathrm{Th}}$          
	 & $I_0$          & $\alpha_I$ & $\beta_\mathcal{L}$ \\
	        &       (nm) &       ($\Omega$) &          ($\upmu$eV)
	        &       ($\upmu$A) &            &           \\
		\hline
	 A & 370/60/20 & 14.0 & 27 & 0.97 & 0.04 & 0.14 \\
	 B & 300/150/20 & 3.6 & 51 & 4.10 & 0.03 & 0.04 \\
	 C & 280/150/20 & 3.0 & 65 & 3.43 & 0.06 & 0.03 \\
 \end{tabular}
 \end{ruledtabular}
 \caption{Summary of key parameters for all the devices.
 Length, width and thickness of each N-wire are reported as $L$, $W$ and $t$
 respectively.
 The Thouless energy ($E_{\mathrm{Th}} = \hbar D / L^2$) 
 has been deduced from the measured normal-state 
 resistance ($R_{\mathrm{n}}$) according to the Einstein relation 
 $D = (\rho_\mathrm{n}\nu_\mathrm{F} e^2)^{-1} $, where $\rho_\mathrm{n}$ is the
 normal-state resistivity of the N-wire and
 $\nu_\mathrm{F} = 1.56 \times 10^{47} \, \mathrm{J}^{-1}\mathrm{m}^{-3}$ 
 is the density of states at the Fermi level for copper.
 $I_0$, $\alpha_I$ and $\beta_\mathcal{L}$ are parameters
 from the RSJ model fitted to 
 $I_{\mathrm{c}}(\Phi)$ experimental data (see Fig.~\ref{fig:icfit}).
 \label{tab:samples}}
\end{table}

The $I$-$V$ characteristics measured for different values of the applied magnetic flux
$\Phi$ allow the investigation of the functional form $I_c(\Phi)$ of the magnetic 
modulation.
Data from three different devices (see Fig.~\ref{fig:icfit}) have been fitted to 
the static zero-temperature resistively shunted junction (RSJ) model\cite{Chesca2004}:
\begin{subequations}
	\label{eqn:modelpre}
	\begin{align}
	i  & = (1-\alpha_I) \sin \delta_1 + (1+\alpha_I) \sin \delta_2 \\
	2 j & = (1-\alpha_I) \sin \delta_1 - (1+\alpha_I) \sin \delta_2 \\
	\delta_2-\delta_1 & = 2\pi \phi + \pi \beta_\mathcal{L} j \label{eqn:deltas} \, ,
\end{align}
\end{subequations}
where $\delta_i$ are the phase differences across the two Josephson junctions,
$\phi = \Phi/\Phi_0$ is the applied magnetic flux normalized to the flux quantum
$\Phi_0 = h/(2e)$,
whereas $i$ and $j$ are supercurrent passing through and circulating in the SQUID,
respectively.
Asymmetries between the two Josephson junctions
are accounted for by the introduction of the $\alpha_I$ parameter.
For fixed applied magnetic flux, the positive and negative critical currents
($I_{\mathrm{c}}^{\pm}$) are defined as proportional to the extremal values of $i$
over all the values of $\delta_1$ and $\delta_2$ that satisfy 
equations~(\ref{eqn:modelpre})
via the coefficient $I_0$, representing the magnitude of the maximum supercurrent
for each weak link of the SQUID:
\begin{equation}
	I_{\mathrm{c}}^{+} = I_0 \max_{\delta_1, \, \delta_2}(i) \quad
	I_{\mathrm{c}}^{-} = I_0 \min_{\delta_1, \, \delta_2}(i) \, .
	\label{eqn:model1}
\end{equation}
The above model includes also a parametric dependence in equation~(\ref{eqn:deltas})
on the inductance $\mathcal{L}$ of the SQUID via the coefficient 
$\beta_\mathcal{L} = 2 \mathcal{L} I_0 / \Phi_0$.
Experimental data show excellent agreement with the theoretical model; fitted
parameters are summarized in table~\ref{tab:samples}.
It is worth emphasizing that albeit the RSJ model was developed for SIS
Josephson junctions it readily applies also to SNS systems provided that their
current-phase relationship is sinusoidal. This is the case for our devices, since
they fall in the \textit{long} junction limit, 
i.e., when the Thouless energy of the junction
is much smaller than the superconducting energy gap 
($\Delta_{\mathrm{BCS}} \approx 0.8 \, \mathrm{meV}$ for bulk vanadium samples).

SQUIDs can be used as magnetometers
in the dissipative regime: by biasing the superconducting ring with a constant current
exceeding the critical current of the interferometer,
changes in magnetic flux can be derived from the corresponding variations in the 
voltage drop developed across the Josephson junctions.
\begin{figure}
 \includegraphics{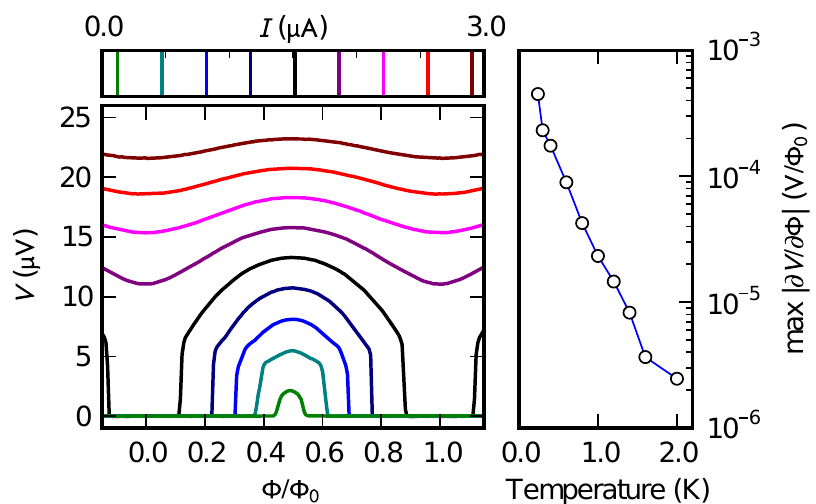}
 \includegraphics{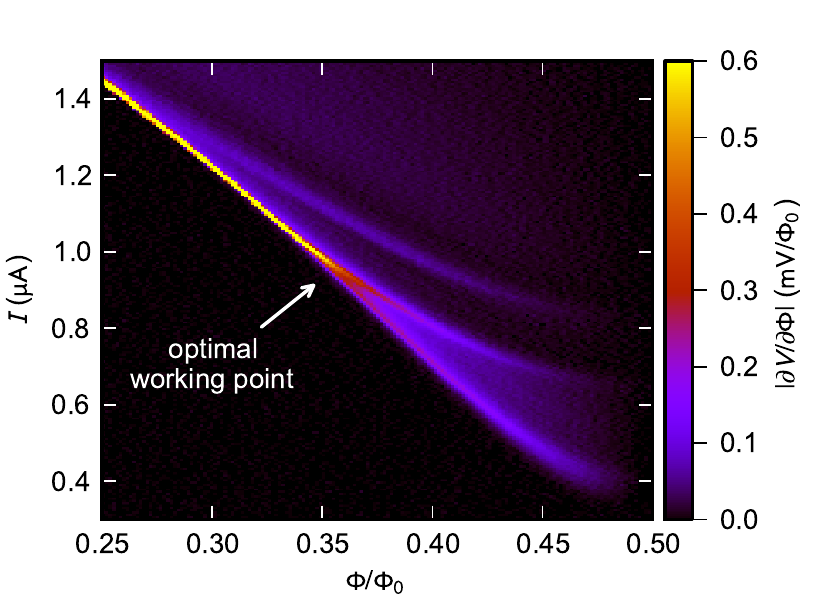}
 \caption{
 \textsc{Top left:}
 Device A $V(\Phi)$ characteristics measured at $240 \, \mathrm{mK}$ for increasing
values of current bias $I$.
 \textsc{Top right:}
 Temperature dependence of the maximum stable value of the flux-to-voltage
 transfer function $\partial V / \partial \Phi$ for device A.  
 The line is a guide to the eye.
 \textsc{Bottom:}
 Map of the transfer function of device A obtained by numerical differentiation
 of $V(\Phi, I)$ data.
 The optimal working point for sensitive magnetometry is indicated by a white arrow
 (see text).
 \label{fig:dvdf}}
\end{figure}

The $V(\Phi)$ characteristics of sample A measured at $240 \, \mathrm{mK}$ are
shown in the top left panel of Fig.~\ref{fig:dvdf} for different values of
the bias current $I$. They are periodic in flux with period $\Phi_0$, and
have an approximate sinusoidal functional form when $I \gg 2 I_0$. In the opposite
limit, the characteristic curves show zero voltage drop $V$ for magnetic flux
values such that $I < I_{\mathrm{c}}(\Phi)$, and finite $V$ values after switching to the
dissipative regime in an interval bounded around $\Phi \approx \Phi_0 (n+1/2)$,
where $n$ is an integer number.
In the switching points themselves the $V(\Phi)$ characteristics display a strongly
 nonlinear behaviour with high values of the flux-to-voltage transfer function 
$\left| \partial V / \partial \Phi \right|$ which, in principle, could allow for highly
sensitive magnetometry. 
However, the switching condition cannot be used as a stable working point 
since the associated dynamic range becomes null as a consequence of the stochastic nature
of the switching.

The transfer function has been obtained 
by numerical differentiation of the $V(\Phi,I)$
characteristics measured in high resolution scans
of the two-dimensional ($\Phi,I$) space. In the resulting map, shown in the bottom panel 
of Fig.~\ref{fig:dvdf}, several ridges are evident from the color plot, 
the most pronounced of which 
corresponds to the aforementioned switching points. As one moves down 
to lower values of the bias current $I$, the profile of the switching ridge
broadens and eventually forks into two different ridges in which the transfer
function reaches values approximately equal to~$ \approx 0.3 \, \mathrm{mV/\Phi_0}$. 

The optimal
working point for maximizing sensitivity corresponds to a bias current
just above the splitting point for the two ridges. In this point, indicated near
the center of the bottom panel of Fig.~\ref{fig:dvdf} by a white arrow, 
the transfer function
obtains values as high as~$\approx 0.45 \, \mathrm{mV/\Phi_0}$ 
and is constant over an effective dynamic
range of approximatively~$10^{-2}\, \Phi_0$ .
The temperature dependence of the maximum (stable) value for the transfer function is
reported in the top right panel of Fig.~\ref{fig:dvdf}, demonstrating the possibility 
of operation 
at temperatures up to~$\approx 2 \, \mathrm{K}$, 
albeit with reduced performance (suppression of approximately one decade per K).

The noise performance of the magnetometers has been characterized by measuring
the power spectral density (PSD) of the signal at the output of the preamplifier
stage. The magnetic flux resolution
of the SQUID is defined as: 
\begin{equation}
	\Phi_{\mathrm{N}} = \frac{\sqrt{S_v}}{
	\left| \partial V / \partial \Phi \right|_{\mathrm{WP}}} \, ,
	\label{eqn:fluxres}
\end{equation}
where $S_v$ is the noise voltage PSD (in $\mathrm{V}^2/\mathrm{Hz}$ units)
and $\left| \partial V / \partial \Phi \right|_{\mathrm{WP}}$ is the flux-to-voltage
transfer function absolute value at the selected working point.
Upon setting the SQUID to its optimal working point, the white noise level was
detected to be~$\sqrt{S_v} = 1.25 \, \mathrm{nV/\sqrt{Hz}}$ at~$1 \, \mathrm{kHz}$, 
which is compatible with the input referred
noise of the preamplifier; this value corresponds to a magnetic flux resolution 
$\Phi_{\mathrm{N}} \approx 2.8 \, \upmu \mathrm{\Phi_0 / \sqrt{Hz}}$ 
at~$1 \, \mathrm{kHz}$.

\begin{figure}[tb]
 \includegraphics{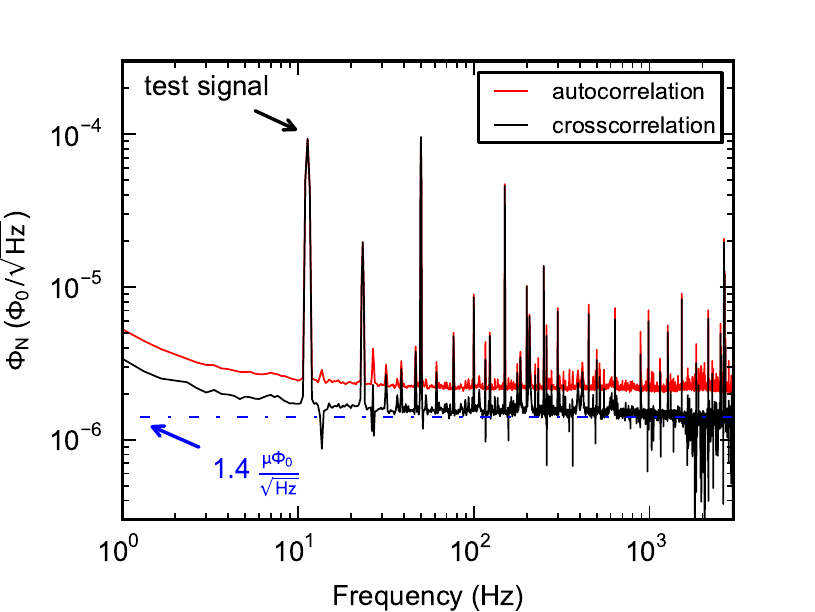}
 \caption{
 Magnetic flux resolution characterization for device A measured 
 at~$240 \, \mathrm{mK}$ and biased at
 the optimal working point for maximum sensitivity. 
 Two independent battery-powered preamplifier units were connected in parallel to the
 SQUID output. The red line represents the flux noise spectrum extracted from
 the autocorrelation of one preamplifier; the black line represents the
 crosscorrelated flux noise between the two preamplifier outputs.
 A small ($\approx 10^{-4}$ $\Phi_0$) applied magnetic flux test signal 
 appears as a peak in the spectra at~$11.7\, \mathrm{Hz}$.
 \label{fig:noise}}
\end{figure}

To test whether the noise limit originates from the preamplification stage itself, 
two independent battery powered LI-75A units were connected in parallel to the
SQUID output. The autocorrelated PSD from one preamplifier
and the crosscorrelated spectral density between the two preamplifiers have been
extracted and compared.
The corresponding magnetic flux resolution spectra are
presented in Fig.~\ref{fig:noise}. 
The autocorrelated spectrum shows the aforementioned 
$2.8 \, \upmu \mathrm{\Phi_0 / \sqrt{Hz}}$ resolution level, whereas the crosscorrelated
one (the blue dashed-dotted line in Fig.~\ref{fig:noise})
reaches a value of~$1.4 \, \upmu \mathrm{\Phi_0 / \sqrt{Hz}}$ 
at~$1 \, \mathrm{kHz}$, thus demonstrating
that the magnetic flux resolution for the SQUIDs is here limited by the 
room-temperature preamplification stage.
We stress that the reported magnetic flux sensitivity levels have been measured 
without the aid of sophisticated electronics or advanced readout schemes, 
and directly follow from the intrinsic voltage response properties of the SNS weak links.

In summary, we presented the fabrication and characterization of DC microSQUIDs based
on V/Cu/V SNS nanojunctions, which offer technological advantages such as long term 
stability, reduced aging, as well as fine geometrical control over the 
transport properties of the Josephson weak links.
Magnetic flux resolution figures 
($\Phi_{\mathrm{N}} < 3 \, \upmu \Phi_0 / \sqrt{\mathrm{Hz}}$), 
already on par with state-of-the-art
commercial DC SQUIDs systems based on SIS technology operating at liquid
He temperature, 
can be further improved by the adoption of
advanced SQUID readout techniques\cite{Drung2004} 
such as the addition of passive and active
flux feedback schemes and cryogenic preamplification stages.
From a more fundamental point of view, the devices herein presented implement
three-terminal Andreev interferometers whose non-trivial dynamics emerge
reproducibly in their phase-dependent transport properties, here exploited
for achieving optimally stable working points for sensitive magnetometry.
This phenomenology, not present in conventional SIS systems, is a typical fingerprint of 
the rich and complex physics underlying the proximity effect\cite{Cuevas2007a}.

\begin{acknowledgments}
The authors acknowledge the Italian Ministry of Defense through the PNRM project
``TERASUPER'', the Marie Curie Initial Training Action (ITN) Q-NET 264034 for partial
financial support. 
C.A. thanks the Tuscany Region for funding his fellowship via the 
CNR joint project ``PROXMAG''.
A.R. thanks Fondazione Tronchetti Provera for funding his Ph.D. scholarship
in Scuola Normale Superiore.
\end{acknowledgments}

\bibliography{paper}
\bibliographystyle{rsc}

\end{document}